\documentstyle[pra,twocolumn,aps]{revtex}
\begin{document}
\draft
\title{
Quantum Langevin equations for semiconductor light-emitting devices
\\
and the photon statistics at a low-injection level}
\author{
Hiroshi Fujisaki \cite{byline1} and Akira Shimizu \cite{byline2}}
\address{
 Institute of Physics, University of Tokyo,
3-8-1 Komaba, Tokyo 153, Japan
}
\date{\today}
\maketitle
\begin{abstract}
From the microscopic quantum Langevin equations (QLEs)
we derive 
the effective semiconductor QLEs and the associated noise correlations 
which are valid at a low-injection level and in real devices.
Applying the semiconductor QLEs to 
semiconductor light-emitting devices (LEDs),
we obtain a new formula for the Fano factor of photons which 
gives the photon-number statistics as a function of the pump statistics
and several parameters of LEDs. 
Key ingredients are non-radiative processes, 
carrier-number dependence of the radiative and non-radiative lifetimes,
and multimodeness of LEDs.
The formula is applicable to the actual cases where the quantum 
efficiency
$\eta$ differs from the differential quantum efficiency $\eta_{\rm d}$, 
whereas previous theories implicitly assumed $\eta = \eta_{\rm d}$.
It is also applicable to the cases 
when photons in each mode of the cavity are emitted and/or detected 
inhomogeneously. 
When $\eta_{\rm d} < \eta$ at a running point,
in particular, our formula predicts 
that even a Poissonian pump can produce sub-Poissonian light.
This mechanism for generation of sub-Poissonian light 
is completely different from those of previous theories, which assumed
sub-Poissonian statistics for the current injected into the 
active layers of LEDs.
Our results agree with recent experiments.
We also discuss frequency dependence of the photon statistics.

\end{abstract}
\section{Introduction}

There have been many researches
 on quantum noise in light-emitting devices (LEDs)
since the celebrated work of Shawllow and Townes \cite{ST}.  
We here consider the quantum noise in 
semiconductor LEDs.
This subject was first studied by Haug and Haken \cite{HH},
who derived from a microscopic model useful formulae
for the first and second optical coherence of LEDs.

Recently, 
much attention has been paid to sub-Poissonian light (SPL),
which has the intensity fluctuations lower 
than the standard quantum limit \cite{Gar2},
of various LEDs
\cite{YM,YL,TS,Bj,TRS,HK,SHKY,HSAK,SA,GKBR,Mar}.
The mechanism for generation of SPL 
in LEDs is quite different from that of 
squeezed light in nonlinear crystals \cite{Sl}.
The latter mechanism is well understood
as a Bogoliuvov transformation of a coherent state 
into a squeezed state of light \cite{MS}.
The former mechanism, on the other hand, is often described by
the Langevin theory of laser \cite{Gar2,Loui,SSL,Haken,WM,CKS}. 
Previous studies on SPL in LEDs assumed that only a 
single mode or a few modes of photons are excited
\cite{YM,TS,Mar}.
When the injection level is lowered, however,
many modes of photons become relevant, 
and we must consider all of them.
Simple theories for such a case were reported in Refs.\cite{YL,Bj}. 
However, they are too simplified so that they cannot explain  
recent experiments by Hirano, Kuga (HK) and coworkers \cite{HK,SHKY,HSAK}, 
who demonstrated that the experimental results at a low-injection level (LIL) 
disagree with the predictions of the simplified theories.
The disagreement appears when the quantum efficiency $\eta$ differs from
the differential quantum efficiency $\eta_{\rm d}$.
HK \cite{HK} suggested that non-radiative processes might be 
responsible for the disagreement.

In this paper, to resolve the discrepancy between the previous theories 
and the experiments,
we theoretically investigate the quantum noise in LEDs 
{\it at a low-injection level} \cite{FS2}.
To this end, we extend the Langevin-equation method of 
Chow, Koch and Sargent \cite{CKS} to treat 
the case where many photon modes are excited in an LED.
From the microscopic quantum Langevin equations (QLEs)
and the associated noise correlations,
we derive
the semiconductor QLEs and the noise correlations
at the LIL.
An important assumption is that
the photon-absorption and emission rates are much smaller than
the photon-escape rate from a ``cavity''.
\footnote{
The validity of this assumption with the meaning of the ``cavity''
will be explained in appendix A.}
In the experiment \cite{SA}, 
it has been reported that $\eta_{\rm d}/\eta > 2$ in low-injection 
regions, whereas
$\eta \simeq \eta_{\rm d}$ in high-injection regions. 
The difference between $\eta$ and $\eta_{\rm d}$ is, in our theory, 
attributed to 
non-radiative processes and carrier-number dependence of lifetimes.
They always exist in real LEDs, and
become particularly important in low-injection regions.
In contrast,
we can easily show that 
the previous simplified theories \cite{YM,YL,TS,Bj}
always give $\eta = \eta_{\rm d}$.
Hence our theory is a minimal one to simulate real LEDs.
A new formula for the photon Fano factor \cite{TS} is derived.
It gives the photon-number statistics as a function of the pump statistics, 
measuring frequency, $\eta$, $\eta_{\rm d}$ and some factors arising from 
{\it multimodeness} of LEDs.
Simplified formulae are derived for
homogeneous cases, in which photons in each mode of the cavity 
are emitted and detected homogeneously, 
and for inhomogeneous cases, in which
they are emitted and/or detected inhomogeneously.
Using these formulae, we discuss 
the condition for generation of SPL in an LED.
Our results agree with the experimental results.

The paper is organized as follows. 
In Section 2,
we derive
the semiconductor QLEs and the noise correlations
at the LIL.
In Section 3, the semiconductor QLEs are used to 
derive a new formula for the photon Fano factor.
In Section 4, we examine the formula in various cases, and
compare our theory with recent experiments \cite{HK,SHKY,HSAK,SA}.
In Section 5, we summarize the paper.

\section{Derivation of quantum Langevin equations and the noise 
correlations 
at a low-injection level}

\subsection{Microscopic Langevin equations of an LED}

Chow {\it et al}. \cite{CKS} discussed the case where
a {\it single mode} is excited among many-photon modes of a cavity. 
Since we are mainly interested in 
the photon statistics of LEDs at the LIL,
we extend their method to treat {\it many modes} of photons.
The total Hamiltonian ${\cal H}_{\rm tot}$ 
(which describes multimode photons
in the cavity and carriers in the active layer of an LED) 
is written as  
\begin{eqnarray}
\label{Ham}
\nonumber
{\cal H}_{\rm tot}
&=&
{\cal H}_{\rm multi-ph}+{\cal H}_{\rm carrier}+{\cal H}_{\rm dipole}
\\
&&+
{\cal H}_{\rm many-body}+{\cal H}_{\rm baths}+{\cal H}_{\rm baths-sys},   
\\
 {\cal H}_{\rm multi-ph}
  &=&     
\sum_l \hbar \nu_l a^{\dagger}_l a_l,
\\
{\cal H}_{\rm carrier}
&=&
\sum_{\bf k} \left[ \left( \epsilon^0_{\rm g}+\frac{\hbar^2 k^2}{2 m_{\rm e}} 
\right) 
c^{\dagger}_{\bf k} c_{\bf k}
     +\frac{\hbar^2 k^2}{2 m_{\rm h}} d^{\dagger}_{\bf -k} d_{\bf -k} \right],
 \\
 {\cal H}_{\rm dipole}
 &=&
 \sum_{l,\bf k} 
\hbar (g^{0}_{l,\bf k} d^{\dagger}_{-\bf k} c^{\dagger}_{\bf k} a_l  + 
{\rm h.c.}),
\end{eqnarray}
where ${\cal H}_{\rm multi-ph}$ is the Hamiltonian of the 
multimode photons, 
$a_l$ is the annihilation operator for the photons in mode $l$,
and $\nu_l$ is the field oscillation frequency in mode $l$. 
The Hamiltonian of  
the electrons and holes in the active layer is ${\cal H}_{\rm carrier}$. 
The annihilation operators of 
the electron and hole of wavevector $\bf k$ are 
$c_{\bf k}$ and $d_{-\bf k}$, respectively, 
$m_{\rm e}$ and $m_{\rm h}$ are the electron and hole effective masses,
respectively,
and $\epsilon^0_{\rm g}$ is the bare band-gap energy.
The interaction among the carriers and the photons is represented 
by ${\cal H}_{\rm dipole}$ in the dipole approximation,
with $g^0_{l,\bf k}$ being the bare coupling constants,
and {\rm h.c.} means Hermite conjugate.
Note that we here consider only a direct radiative transition,
however, in  real devices, there are other processes, 
and we will include them when necessary.
The many-body interaction between the 
carriers is represented by ${\cal H}_{\rm many-body}$, 
the Hamiltonian of baths (or environments) by
${\cal H}_{\rm baths}$,
and the interaction  
between the baths and the system (the carriers and photons) by 
${\cal H}_{\rm baths-sys}$.

To obtain QLEs, we start with making a mean-field approximation for 
${\cal H}_{\rm many-body}$,
and consequently 
$\epsilon^0_{\rm g}$ and $g^0_{{\bf k},l}$ 
are renormalized (See, {\it e.g.}, Chap. 4 of Ref. \cite{CKS}).
The renormalized parameters are denoted by 
$\epsilon_{\rm g}$ and $g_{l,\bf k}$.

We then eliminate ${\cal H}_{\rm baths}+{\cal H}_{\rm baths-sys}$
by using the Markov approximation \cite{Gar2,MS,Loui,SSL,Haken,WM,CKS}. 
As a result, the fluctuation and dissipation terms appear in 
the equations of motion.
The generalized Einstein relation \cite{MS,Loui,SSL,WM,CKS} 
gives the relation between 
the fluctuation and dissipation terms as follows,
\begin{equation}
\label{Ein}
2D_{\mu \nu}=
\frac{d}{dt}\langle A_\mu A_\nu \rangle 
-\langle D_\mu A_\nu \rangle -\langle A_\mu D_\nu \rangle,
\end{equation}
where $2D_{\mu \nu}$ is a diffusion coefficient, 
$A_\mu$ a system variable,
$D_\mu$ a dissipation term in Langevin equations,
i.e., $\dot{A}_{\mu}=D_{\mu}+F_{\mu}, 
\langle F_\mu(t)F_\nu(t')\rangle =2D_{\mu \nu}\delta(t-t')$.
The brackets mean the ensemble average for fluctuations.

We finally obtain the following
microscopic QLEs, which describe LEDs in a microscopic scale, 
for the dipole operator $\sigma_{\bf k} = d_{-\bf k} c_{\bf k} e^{i \nu_l t}$, 
for the electric field operator $A_l(t)=a_l(t) e^{i \nu_l t}$,
and for the electron occupation probability in $\bf k$ space 
$n_{\rm e \bf k} = c^{\dagger}_{\bf k} c_{\bf k}$,
\begin{eqnarray}
\nonumber
\frac{d}{dt} \sigma_{\bf k} 
&=&
 -(\gamma+i \omega_{\bf k}-i \nu_l) \sigma_{\bf k} 
\\
\label{Langevin_for_sig}
&&
+i \sum_{l'} g_{l',\bf k} A_{l'}(n_{\rm e \bf k}+n_{\rm h \bf k}-1)
+F_{\sigma \bf k},
\\
\label{Langevin_for_A}
\frac{d}{dt} A_l 
&=& -[\frac{\kappa_l^0}{2} +i(\Omega_l-\nu_l)]A_l
-i \sum_{\bf k} g^*_{l,\bf k} \sigma_{\bf k} +F_l,
\\
\nonumber
\frac{d}{dt}n_{\rm e \bf k}
&=&
P_{\rm e \bf k}(1-n_{\rm e \bf k})
-\frac{n_{\rm e \bf k}}{\tau_{\rm nr}}
\\
\label{Langevin_for_carrier}
&&
+\sum_l 
(i g^*_{l, \bf k}A^{\dagger}_l \sigma_{\bf k} +{\rm h.c.}) 
+F_{\rm e \bf k},
\end{eqnarray}
where 
$\gamma$ is the dipole decay (dephasing) rate,
$\hbar \omega_{\bf k} \equiv \epsilon_{\rm g}
+\hbar^2 k^2/2 m_{\rm e}+\hbar^2 k^2/2 m_{\rm h}$ is the transition energy,
and $F_{\sigma \bf k}$ is the fluctuation operator for the dipole.
The hole occupation probability in $\bf k$ space is
$n_{\rm h \bf k} \equiv d^{\dagger}_{-\bf k} d_{-\bf k}$.
The photon escape rate from the cavity is 
$\kappa_l^0=\nu_l/Q_l$ where
$Q_l$ is the Q factor of the cavity, $\Omega_l$ is the passive-cavity 
frequency, and
$F_l$ is the fluctuation operator for the electric field.
The pump rate due to a current injection or optical pumping is
$P_{\rm e \bf k}(1-n_{\rm e \bf k})$
where the factor $(1-n_{\rm e \bf k})$ represents 
the pump blocking \cite{CKS}.
Note that a lifetime of non-radiative decay 
$\tau_{\rm nr}$ has been introduced in Eq.\ (\ref{Langevin_for_carrier}).  
As discussed later, the existence of non-radiative processes
is, in our model, a necessary condition for the difference between 
the quantum efficiency and the differential quantum efficiency to occur.
The other condition is that the lifetime of radiative processes or
that of non-radiative ones varies with $n_{\rm c}$. 
[See Eqs.\ (\ref{qe})-(\ref{def2}).]
Non-radiative processes might be modelled 
by capture of carriers at a trapping level in an LED, 
however, we only need that $\tau_{\rm nr}$ is an implicit function of 
$n_{\rm c}$.
These terms of pump and non-radiative decay
have been phenomenologically introduced.
The fluctuation operator for the electron number is 
$F_{\rm e \bf k}$.

\subsection{Adiabatic approximation }
To derive more useful forms for later discussion,
we use the adiabatic approximation \cite{Haken,Gar},
and approximate the solution of Eq.\ (\ref{Langevin_for_sig}) by
\begin{equation}
\label{sig2}
\sigma_{\bf k} 
\simeq 
\frac{i \sum_{l'} g_{l',\bf k} A_{l'} (n_{\rm e \bf k}+n_{\rm h \bf k}-1)+F_{\sigma \bf k}}
{\gamma+i \omega_{\bf k}-i \nu_l},
\end{equation}

Substituting Eq.\ (\ref{sig2}) into (\ref{Langevin_for_A}), we find
\begin{eqnarray}
\label{A}
\nonumber
\dot A_l
 &=& 
 -[\kappa_l^0/2 +i(\Omega_l-\nu_l)]A_l
\\
&&
+\sum_{l'} G_{ll'} A_{l'}
+F_l+F_{\sigma,l},
\end{eqnarray}
where $G_{ll'}$ is a ``gain matrix";
\begin{equation}
\label{gainmatrix}
G_{ll'} \equiv 
\sum_{\bf k} g^*_{l,\bf k} g_{l',\bf k} {\cal D}_{l,\bf k}
(n_{\rm e \bf k}+n_{\rm h \bf k}-1),
\end{equation}
where ${\cal D}_{l,\bf k}$ is a complex Lorentzian; 
$
{\cal D}_{l,\bf k} \equiv \frac{1}{\gamma+i (\omega_{\bf k}- \nu_l)}.
$
A new fluctuation operator $F_{\sigma,l}(t)$ has been defined by
\begin{equation}
F_{\sigma,l}
\equiv
-i \sum_{\bf k} g^*_{l,\bf k}{\cal D}_{l,\bf k}F_{\sigma \bf k},
\end{equation} 
which is associated with the coupling between the carriers and photons.

From Eq.\ (\ref{A}), we also find the QLE for 
the photon-number operator $n_l \equiv A^{\dagger}_l A_l$,
\begin{eqnarray}
\label{ln}
\nonumber
\frac{d}{dt}n_l 
&=& - \kappa_l^0 n_l+ \sum_{l'} [G_{ll'} A^{\dagger}_l A_{l'}+ {\rm h.c.}]
\\
&&
+[ (F^{\dagger}_{\sigma,l} +F^{\dagger}_l)A_l+{\rm h.c.}],
\end{eqnarray}

Substituting Eq.\ (\ref{sig2}) into (\ref{Langevin_for_carrier}), 
we also find the QLE for the total electron number 
$n_{\rm c} \equiv \sum_{\bf k} n_{\rm e \bf k}$, 
\begin{eqnarray}
\label{lc}
\nonumber
\frac{d}{dt} n_{\rm c}
&=&
\sum_{\bf k} P_{\rm e \bf k}(1-n_{\rm e \bf k})
 -\frac{n_{\rm c}}{\tau_{\rm nr}}
\\
\nonumber
&&
-\sum_{l,l'}
[G_{ll'} A^{\dagger}_l A_{l'}+ {\rm h.c.}]
\\
&&
+\sum_{\bf k} F_{{\rm e} \bf k}
-\sum_l [A^{\dagger}_l F_{\sigma,l} +{\rm h.c.}],
\end{eqnarray}

\subsection{Microscopic noise correlations of an LED}

To discuss the statistical properties of light emitted from LEDs,
we must determine the noise correlations.
Hereafter we assume that 
the correlations between different modes of the photons and 
those between different wavenumbers of carriers 
can be neglected \cite{Haken}, {\it i.e.},
\begin{eqnarray}
\label{mode}
\langle A^{\dagger}_l (t) A_{l'}(t) \rangle 
&\simeq& \langle n_l \rangle \delta_{ll'},
\\
\langle \sigma^{\dagger}_{\bf k}(t) \sigma_{\bf k'}(t) \rangle 
&\simeq& \langle n_{\rm e \bf k}n_{\rm h \bf k} \rangle \delta_{\bf k,k'},
\\
\langle \sigma_{\bf k}(t) \sigma^{\dagger}_{\bf k'}(t) \rangle
&\simeq& \langle (1-n_{\rm e \bf k})(1-n_{\rm h \bf k}) \rangle \delta_{\bf k,k'},
\\
\langle n_{{\rm e}\bf k}(t) n_{{\rm e}\bf k'}(t) \rangle
&\simeq& \langle n_{{\rm e}\bf k} \rangle \delta_{\bf k,k'}.
\end{eqnarray}
We can then calculate
the noise correlations, which are consistent with the QLEs 
(\ref{Langevin_for_sig})-(\ref{Langevin_for_carrier}),
using (\ref{Ein});
\begin{eqnarray}
\label{csig}
\langle F_{\sigma \bf k}^{\dagger}(t) F_{\sigma \bf k'}(t') \rangle
&\simeq&
2 \gamma \langle n_{\rm e \bf k}n_{\rm h \bf k} \rangle \delta_{\bf k,k'}\delta(t-t'),
\\
\label{csig2}
\langle F_{\sigma \bf k}(t) F^{\dagger}_{\sigma \bf k'}(t') \rangle
&\simeq&
2 \gamma \langle (1-n_{\rm e \bf k})(1-n_{\rm h \bf k}) \rangle 
\nonumber
\\
&&\delta_{\bf k,k'}\delta(t-t'),
\\
\label{cf}
\langle F_l^{\dagger}(t) F_{l'}(t') \rangle
&\simeq& \kappa_l^0 \overline n(\nu_l) \delta_{ll'} \delta(t-t'),
\\
\label{cr2}
\langle F_l(t) F_{l'}^{\dagger}(t')  \rangle
&\simeq& \kappa_l^0 \left [ \overline n(\nu_l)+1 \right ] 
\delta_{ll'} \delta(t-t'),
\\
\nonumber
\langle F_{{\rm e} \bf k}(t) F_{{\rm e} \bf k'}(t') \rangle 
&=&
\left [
\langle P_{\rm e \bf k}(1-n_{\rm e \bf k}) \rangle
\label{ce}
+\langle n_{\rm e \bf k}/\tau_{\rm nr} \rangle 
\right ] 
\\
&&
\delta_{\bf k,k'} \delta(t-t'),
\end{eqnarray}
where 
$\overline n (\nu_l)$ is the number of thermal photons in mode $l$ in the cavity.
We have used,
in Eqs.\ (\ref{csig}) and (\ref{csig2}),
the quasiequilibrium conditions \cite{CKS},
\begin{eqnarray}
2 \gamma \langle n_{\rm e \bf k}n_{\rm h \bf k} \rangle
&\gg& \frac{d}{dt}  \langle n_{\rm e \bf k}n_{\rm h \bf k}\rangle,
\\ 
2 \gamma \langle (1-n_{\rm e \bf k})(1-n_{\rm h \bf k}) \rangle 
&\gg& \frac{d}{dt}\langle (1-n_{\rm e \bf k})(1-n_{\rm h \bf k}) \rangle.
\end{eqnarray}

We can also calculate the correlations of $F_{\sigma,l}$;
\begin{eqnarray}
\label{cf1}
\langle F^{\dagger}_{\sigma,l}(t) F_{\sigma,l} (t') \rangle 
&=&
\langle R_{{\rm sp},l} \rangle \delta(t-t'),
\\
\label{cf2}
\langle F_{\sigma,l} (t) F^{\dagger}_{\sigma,l}(t') \rangle 
&=&
\langle R_{{\rm abs},l} \rangle \delta(t-t'),
\end{eqnarray}
where
$R_{{\rm sp},l}$ ($R_{{\rm abs},l}$) denotes
the spontaneous emission rate
into photon mode $l$ (the absorption rate of photon mode $l$);
\begin{eqnarray}
\label{Rsp}
R_{{\rm sp},l} 
&\equiv&
\frac{n_{\rm c}}{\tau_{{\rm r},l}} 
\equiv 
\frac{2}{\gamma} 
\sum_{\bf k} 
|g_{l,\bf k}|^2 {\cal L}_{l, \bf k} n_{\rm e \bf k}n_{\rm h \bf k},
\\
\label{Rabs}
R_{{\rm abs},l} 
&\equiv&
\frac{2}{\gamma} 
\sum_{\bf k} 
|g_{l,\bf k}|^2 {\cal L}_{l,\bf k} (1-n_{\rm e \bf k})(1-n_{\rm h \bf k}).
\end{eqnarray}
We have used Eqs.\ (\ref{csig}) and (\ref{csig2}),
and defined 
$\tau_{{\rm r},l}$ as a radiative lifetime of the carriers into mode $l$.
The (dimensionless) 
Lorentzian line-shape function ${\cal L}_{l,\bf k}$ is
$
{\cal L}_{l,\bf k} \equiv \frac{\gamma^2}{\gamma^2+(\omega_{\bf k}
-\nu_l)^2}.
$

At this stage, we have the following QLEs,
\begin{eqnarray}
\label{ln2}
\frac{d}{dt}n_l 
&=& - \kappa_l n_l+ 
[ (F^{\dagger}_{\sigma,l} +F^{\dagger}_l)A_l+{\rm h.c.}],\\
\nonumber
\frac{d}{dt} n_{\rm c}
&=&
\sum_{\bf k} P_{\rm e \bf k}(1-n_{\rm e \bf k}) 
-\frac{n_{\rm c}}{\tau_{\rm nr}}
\\
\nonumber
&&
-\sum_{l}(R_{{\rm sp},l}-R_{{\rm abs},l})n_l \\
&&+\sum_{\bf k} F_{{\rm e} \bf k}
-\sum_l [A^{\dagger}_l F_{\sigma,l} +{\rm h.c.}],
\label{lnc2}
\end{eqnarray}
where we have used
\begin{equation}
G_{ll}+G^*_{ll}
=R_{{\rm sp},l}-R_{{\rm abs},l},
\end{equation}
and defined the renormalized photon escape rate
\begin{equation}
\kappa_l
=\kappa_l^0-R_{{\rm sp},l}+R_{{\rm abs},l}.
\end{equation}



\subsection{Semiconductor Langevin equations and the noise correlations 
for an LED at a low-injection level}
In what follows, we assume
\footnote{
The following inequality, at first sight, might look rather strange.
We therefore explain 
its validity {\it at the LIL} in the appendix A.}
\begin{equation}
\label{condition_for_low_injection}
R_{{\rm sp},l}, R_{{\rm abs},l} \ll \kappa_l^0.
\end{equation}
From Eqs.\ (\ref{ln2}) and (\ref{condition_for_low_injection}),
the QLE for $n_l$ at the LIL becomes
\begin{equation}
\label{lnt}
\nonumber
\frac{d}{dt}n_l 
= -\kappa_l^0 n_l + R_{{\rm sp},l} + F_{n,l},
\end{equation}
where we have defined 
a new fluctuation operator;
\begin{equation}
F_{n,l} 
\equiv
 [ (F^{\dagger}_{\sigma,l} +F^{\dagger}_l)A_l +{\rm h.c.}]
 -(\kappa_l^0 \overline{n}(\nu_l)+R_{{\rm sp},l}).
\end{equation}
The right-hand side is averaged to be zero because the noises, $F_{n,l}$ and 
$F_{\sigma,l}$,
have the Markovian property:
$\langle F^{\dagger}_l A_l +{\rm h.c.} \rangle 
=\kappa_l^0 \overline{n}(\nu_l)$
and $\langle F^{\dagger}_{\sigma,l} A_l +{\rm h.c.} \rangle = R_{{\rm sp},
l}$.

From Eqs.\ (\ref{condition_for_low_injection}) and (\ref{lnt}),
the steady-state value of the photon number becomes
\begin{equation}
\label{n_l<<1}
(n_l)_{\rm s.s.} \simeq 
R_{{\rm sp},l}/\kappa_l^0 \ll 1,
\end{equation}
i.e., the photon number in each mode is quite small.

On the other hand, from Eqs.\ (\ref{lnc2}), 
(\ref{condition_for_low_injection}), and (\ref{n_l<<1}),
the QLE for $n_{\rm c}$ 
at the LIL becomes
\begin{equation}
\label{lnclow}
\frac{d}{dt} n_{\rm c}
=
P-\frac{n_{\rm c}}{\tau_{\rm r}} -\frac{n_{\rm c}}{\tau_{{\rm nr}}} 
+F_{\rm c},
\end{equation}
where
$P \equiv 
\sum_{\bf k} P_{e \bf k}(1-n_{e \bf k})$ is the total pump rate, 
and
\begin{equation}
\label{tau}
\frac{1}{\tau_{\rm r}} 
\equiv \frac{\sum_l R_{{\rm sp},l}}{n_{\rm c}} 
\equiv \sum_l \frac{1}{\tau_{{\rm r},l}},
\end{equation}
is the radiative decay rate 
($\tau_{\rm r}$ is the radiative lifetime of carriers).
The fluctuation operator for the total electron 
number $F_{\rm c}$ is denoted by,
\begin{equation}
F_{\rm c} 
\equiv
\sum_l
\left [
-(A^{\dagger}_l F_{\sigma,l} +{\rm h.c.}) +R_{{\rm sp},l}
\right ]
+\sum_{\bf k} F_{{\rm e} \bf k}.
\end{equation}
Note that 
the lifetimes ($\tau_{{\rm r},l}$ and $\tau_{\rm r}$)
are carrier-number 
$n_{\rm c}$ dependent because $R_{{\rm sp},l}$ 
is an implicit function of $n_{\rm c}$. 

We thus obtain the final forms of the semiconductor QLEs 
at the LIL as follows,
\begin{eqnarray}
\label{l1}
\frac{d}{dt} n_{\rm c} 
&=& P - \frac{n_{\rm c}}{\tau_{\rm r}} -\frac{n_{\rm c}}{\tau_{{\rm nr}}} 
+\Gamma_{\rm P}+\Gamma_{\rm r}+\Gamma_{\rm nr},
\\
\label{l2}
\frac{d}{dt} n_l
&=& -\kappa_l^0 n_l +\frac{n_{\rm c}}{\tau_{{\rm r},l}} 
+F_{\kappa,l}+F_{{\rm r},l},
\end{eqnarray}
where the noise operators, $F_{n,l}, F_{\rm c}$, 
have been divided into new ones, 
$\Gamma_{\rm r}, F_{{\rm r},l}, \Gamma_{\rm nr}, \Gamma_{\rm P}, 
F_{\kappa,l}$.
They are given by
\begin{eqnarray}
\Gamma_{\rm r}
&\equiv&
\sum_l 
\left [
-(A^{\dagger}_l F_{\sigma,l}+{\rm h.c.})+R_{{\rm sp},l}
\right],
\\
F_{{\rm r},l}
&\equiv&
A^{\dagger}_l F_{\sigma,l}+{\rm h.c.}- R_{{\rm sp},l},
\\
\Gamma_{\rm nr}+\Gamma_{\rm P}
&\equiv&
\sum_{\bf k} 
F_{{\rm e} \bf k},
\\
F_{\kappa,l}
&\simeq&
A^{\dagger}_l F_l + F^{\dagger}_l A_l,
\end{eqnarray}
where we have neglected the number of thermal photons, 
$\overline{n}(\nu_l)$,
since this is the case in usual experiments.
These noise operators
($\Gamma_{\rm r}, F_{{\rm r},l}, \Gamma_{\rm nr},
\Gamma_{\rm P}, F_{\kappa,l}$) 
are associated with the radiative decay of carriers,
the conversion from carriers to photons, the non-radiative decay of 
carriers,
the (intrinsic) pump fluctuation, and the photon escape from the cavity, 
respectively.
The mean values of the noises are defined to be zero, of course.

From above the noise correlation for $\Gamma_{\rm r}$
can be calculated as,
\begin{eqnarray}
\nonumber
\lefteqn{
\langle \Gamma_{\rm r}(t) \Gamma_{\rm r}(t') \rangle
}
\\
\nonumber
&\simeq&
\sum_l
\left[
\langle A^{\dagger}_{{\rm c},l}(t) A_{{\rm c},l}(t') \rangle 
\langle F_{\sigma,l}(t) F^{\dagger}_{\sigma,l}(t') \rangle 
\right .
\\
\nonumber
&&
+
\left .
\langle A_{{\rm c},l}(t) A^{\dagger}_{{\rm c},l}(t') \rangle 
\langle F^{\dagger}_{\sigma,l}(t) F_{\sigma,l}(t') \rangle 
\right]
\\
\nonumber
&=&
\sum_l
[
\langle R_{{\rm sp},l}+R_{{\rm abs},l} \rangle \langle n_l \rangle 
+\langle R_{{\rm sp},l} \rangle
]
\delta(t-t')
\\
\label{cgr}
&\simeq&
\sum_l \langle R_{{\rm sp},l} \rangle \delta(t-t')
 = \left < \frac{n_{\rm c}}{\tau_{\rm r}} \right >
\delta(t-t'),
\end{eqnarray}
where,
as done in Eq.\ (\ref{mode}), 
we have neglected,
in the second and third lines, 
the correlations between different modes.
We also have assumed the following,
\begin{eqnarray}
\label{sagyou}
\nonumber
\Gamma_{\rm r}
&\simeq&
\sum_l
\left [
-(A^{\dagger}_{{\rm c},l} F_{\sigma,l} +{\rm h.c.})
\right],
\end{eqnarray}
where $A^{\dagger}_{{\rm c},l}$ is an operator
uncorrelated with $F_{\sigma,l}$ \cite{F2}.
Thus the property of fluctuations, $\langle \Gamma_{\rm r} \rangle =0$, 
still holds.
Equations (\ref{cf1}) and (\ref{cf2}) have also been used   
in the fourth line of Eq.\ (\ref{cgr}), and
Eqs.\ (\ref{condition_for_low_injection}),(\ref{n_l<<1}), and (\ref{tau})
in the fifth line of Eq.\ (\ref{cgr}).

Similarly we obtain
\begin{eqnarray}
\langle \Gamma_{\rm r}(t) F_{{\rm r},l}(t') \rangle 
&=&
-\left < \frac{n_{\rm c}}{\tau_{{\rm r},l}} \right > \delta(t-t'),
\\
\langle F_{{\rm r},l}(t) F_{{\rm r},l'}(t') \rangle
&=&
\left < \frac{n_{\rm c}}{\tau_{{\rm r},l}} \right > \delta_{ll'}\delta(t-t'),
\\
\langle \Gamma_{\rm nr}(t) \Gamma_{\rm nr}(t') \rangle
&=&
\left < \frac{n_{\rm c}}{\tau_{{\rm nr}}} \right > \delta(t-t'),
\\
\langle \Gamma_{\rm P}(t) \Gamma_{\rm P}(t') \rangle 
&=&
\langle P \rangle \delta(t-t'),
\label{cnrp}
\end{eqnarray}
where we have divided the correlation
$\langle \Gamma_{\rm nr}(t) \Gamma_{\rm nr}(t') \rangle
+\langle \Gamma_{\rm P}(t) \Gamma_{\rm P}(t') \rangle$ 
into two parts;
one is the correlation of the non-radiative processes, 
and the other that of the pump.
These semiconductor QLEs, (\ref{l1}) and (\ref{l2}), 
and the noise correlations (\ref{cgr})-(\ref{cnrp})
were just assumed by us \cite{FS2,FS} and Yamanishi and Lee (YL) \cite{YL},
although YL did not include non-radiative processes.

Furthermore, we focus on
the total photon flux, $N$, 
which can be detected at the PD surface.
To this end, we must find two relations: one is the relation between the 
photon number
in the cavity and the photon flux from the cavity. The other is the relation 
between 
the photon flux from the cavity and the photon flux detected at the PD surface.
The input-output formalism \cite{Gar2,WM} gives the first relation as,
\begin{equation}
\label{l3}
V_l = \kappa_l^0 n_l -F_{\kappa,l},
\end{equation}  
where $V_l$ is the photon flux of mode $l$ from the cavity.
We next model the second relation using
the argument of a beam splitter \cite{WM}.
The relation between the mean flux of
$N$ and that of $V_l$
is given by
\begin{equation}
\label{Nmean}
\langle N \rangle =\sum_l \xi_l \langle V_l \rangle 
\equiv \beta_0 \sum_l \langle V_l \rangle
\equiv \beta_0 \langle V \rangle,
\end{equation}
where $\xi_l$ is a ``transmission coefficient" of mode $l$ (see Fig.\ 1),
\begin{figure}
\label{fig1}
\caption{
Definition of the $\xi_l$.
Photons of mode $l$ are emitted into the region between the solid lines.
The photodetector can detect only some region (denoted by
the one between the dashed lines) of the photons
because its surface
area is limited and the quantum efficiency $\eta$ of the detector is 
less than one.
$\xi_l$ is the probability that a photon of mode $l$ is detected by 
the detector.
It increases as the surface area and/or $\eta$ do. 
}
\end{figure}
$\beta_0$ is the transfer efficiency \cite{YL}, 
and  $V \equiv \sum_l V_l$ is the total photon flux from the cavity.
The relation between the flux correlation of $N$ and that of $V_l$  
is given by
\begin{eqnarray}
\nonumber
\langle |\Delta \tilde N(\Omega)|^2 \rangle 
= \sum_l \xi_l (1-\xi_l) \langle \tilde V_l(\Omega) \rangle 
\\
\label{Ncorr}
+\sum_{l,l'} \xi_l \xi_{l'} \langle \Delta \tilde V_l(\Omega) \Delta 
\tilde V_{l'}(\Omega) \rangle,
\end{eqnarray}
where $\tilde N(\Omega)$ and $\tilde V_l(\Omega)$ denote Fourier 
components of $N$ and $V_l$, respectively.

\section{Calculation of the photon Fano factor}

In this section, we calculate the Fano factor of photons\cite{TS} 
which denotes the normalized fluctuation 
of the photon number detected at the PD surface.

Following the standard small-signal analysis used in 
Refs.\cite{YM,YL,Bj,Mar},
we expand the radiative and non-radiative lifetimes,
$\tau_{{\rm r},l}[n_{\rm c}]$, $\tau_{\rm r}[n_{\rm c}]$, and 
$\tau_{{\rm nr}}[n_{\rm c}]$, 
to linear order in 
$\Delta n_{\rm c} \equiv n_{\rm c}-n_{{\rm c}0}$; 
\begin{eqnarray}
\label{trl}
  \tau_{{\rm r},l}[n_{\rm c}]
    &=& (\tau_{{\rm r},l})_0 (1-K_{{\rm r},l} \frac{\Delta {n_{\rm 
c}}}{n_{{\rm c}0}}),
\\
  \label{tr}
  \tau_{\rm r}[n_{\rm c}]
    &=& \tau_{{\rm r}0} (1-K_{\rm r} \frac{\Delta {n_{\rm c}}}{n_{{\rm 
c}0}}),
\\
  \label{tnr}
  \tau_{{\rm nr}}[n_{\rm c}]
    &=& \tau_{{\rm nr}0} (1+K_{{\rm nr}} \frac{\Delta {n_{\rm c}}}{n_{{\rm 
c}0}}),
\\
\frac{1}{\tau_{{\rm r}0}} &\equiv& \sum_l \frac{1}{(\tau_{{\rm r},l})_0},
\\
K_{\rm r} &=& \sum_l \frac{\tau_{{\rm r}0}}{(\tau_{{\rm r},l})_0} K_{{\rm 
r},l},
\end{eqnarray}
where
quantities with the subscript 0 denote the average values
at a running point, $P=P_0$.
Sensitivity of lifetimes to the carrier-number
fluctuations $\Delta n_{\rm c}$ is represented by
$K_{{\rm r},l}, K_{\rm r}$ and $K_{{\rm nr}}$.
Though $K_{\rm r}$ is usually positive, it is
not always the case as we will discuss in the next section.

Linearizing Eqs.\ (\ref{l1}), (\ref{l2}) and (\ref{l3}) 
in terms of
$\Delta n_{\rm c}$,
$\Delta n_l \equiv n_l - (n_l)_0$,
$\Delta P \equiv P - P_0$, and
$\Delta V _l\equiv V_l -(V_l)_0$, 
and substituting Eqs.\ (\ref{trl}) and (\ref{tnr}) into them, we obtain
\begin{eqnarray}
  \label{ll1}
  \frac{d}{dt} \Delta n_{\rm c} &=&
               \Delta P -\frac{\Delta n_{\rm c}}{\tau''}
               +\Gamma_{\rm P}+\Gamma_{\rm r}+\Gamma_{\rm nr},
\\
  \label{ll2}
  \frac{d}{dt} \Delta n_l &=& -\kappa_l^0 \Delta n_l
                       +\frac{\Delta n_{\rm c}}{\tau_{{\rm r},l}'} 
                       +F_{\kappa,l}+F_{{\rm r},l},
\\
  \label{ll3}
  \Delta V_l &=& \kappa_l^0 \Delta n_l-F_{\kappa,l},
\end{eqnarray}
where we have used the following equilibrium conditions:
\begin{eqnarray}
  \label{equi}
\nonumber
  P_{0} &=& n_{{\rm c}0} (\frac{1}{\tau_{{\rm r}0}}+\frac{1}{\tau_{{\rm 
nr}0}}),
\\     
\nonumber
  \sum_l \kappa_l^0 (n_l)_0 &=& \sum_l \frac{n_{{\rm c}0}}{(\tau_{{\rm 
r},l})_0}= \frac{n_{{\rm c}0}}{\tau_{{\rm r}0}}
 \\
  &=&\sum_l V_{l0} = V_0 =\frac{P_0}{1+(\tau_{{\rm r}0}/\tau_{{\rm 
nr}0})},
\end{eqnarray}
and introduced the effective lifetimes defined as
\begin{eqnarray}
  \label{def}
\tau_{{\rm r}}' \equiv \frac{\tau_{{\rm r}0}}{1+K_{\rm r}},
\hspace{3mm} 
 \tau_{{\rm r},l}' &\equiv& \frac{(\tau_{{\rm r},l})_0}{1+K_{{\rm r},l}},
\hspace{3mm} 
  \tau_{{\rm nr}}' \equiv \frac{\tau_{{\rm nr}0}}{1-K_{{\rm nr}}},
\\
\label{cut}
 \frac{1}{\tau''} &\equiv& \frac{1}{\tau_{\rm r}'} + \frac{1}{\tau_{{\rm 
nr}}'}.
\end{eqnarray}
Dropping the noise terms in Eqs.\ (\ref{ll1})-(\ref{ll3}), and
using Eq.\ (\ref{Nmean}), 
we calculate
the quantum efficiency $\eta$ and the differential quantum 
efficiency $\eta_{\rm d}$ as 
\begin{eqnarray}
  \label{qe}
  \eta &\equiv& \frac{N_0}{P_0}
       = \frac{\beta_0/\tau_{{\rm r}0}}{(1/\tau_{{\rm r}0}+1/\tau_{{\rm 
nr}0})}
       =\frac{\beta_0}{1+\epsilon_0},
\\
  \label{dqe}
  \eta_{\rm d} &\equiv& 
        \left. \frac{\Delta \tilde N (\Omega)}
             {\Delta \tilde P (\Omega)} \right |_{\Omega \rightarrow 0}
        =\frac{\beta_0/\tau_{{\rm r}}'}{(1/\tau_{{\rm r}}'+1/\tau_{{\rm 
nr}}')}
        =\frac{\beta_0}{1+\epsilon'},
\end{eqnarray}
where 
\begin{equation}
  \label{def2}
  \epsilon_0 \equiv \frac{\tau_{{\rm r}0}}{\tau_{{\rm nr}0}},
\hspace{5mm}
  \epsilon' \equiv \frac{\tau_{\rm r}'}{\tau_{\rm nr}'}
             =\frac{1-K_{{\rm nr}}}{1+K_{\rm r}} \epsilon_0.
\end{equation}
These efficiencies ($\eta$ and $\eta_{\rm d}$) are illustrated in Fig.\ 2.
\begin{figure}
\caption{
Typical I-L (Injection-Light) characteristics of LEDs
are shown by the solid lines. 
The quantum efficiency $\eta$ is the slope of the straight line
(represented by the dashed line) 
connecting the origin and the running point, whereas 
the differential quantum efficiency $\eta_{\rm d}$ is 
the slope of the tangential line (chain line) at the running point.
}
\label{fig2}
\end{figure}
Note that the difference between $\eta$ and $\eta_{\rm d}$,
 which was measured to be large
({\it e.g.}, $\eta_{\rm d}/\eta > 2$
at the LIL \cite{SA}),
can not be explained by the previous theories
which assumed $\tau_{{\rm nr}0}=\infty$ \cite{YM,YL,TS} 
or $K_{\rm r}=K_{{\rm nr}}=0$ \cite{Bj}.
In this sense, our model is a {\it minimal} one that simulates 
real LEDs.

Hereafter we assume that
$\Omega \ll \kappa_l^0$ ($\sim 10^{12}$Hz),
which is well satisfied in usual experimental conditions.
Then the response of the output photon flux to the modulation of the pump 
is calculated as,
\begin{equation}
\label{mod}
\left | \frac{\Delta \tilde N(\Omega)}{\Delta \tilde P(\Omega)} \right |
=\frac{\eta_{\rm d}}{\sqrt{1+\Omega^2 \tau''^2}}.
\end{equation}
Hence $1/\tau''$ is understood as a cutoff frequency.
It is seen that the non-radiative processes push the cutoff 
to the higher frequency [See Eq.\ (\ref{cut})].

On the other hand,
the photon flux correlation between mode $l$ and $m$ 
is calculated using Eqs.\ (\ref{ll1})-(\ref{ll3}) as 
\begin{eqnarray}
\label{out}
& & 
\nonumber
\langle \Delta \tilde V_l^{*}(\Omega) \Delta \tilde V_m(\Omega) \rangle
\\
\nonumber
&=& 
\frac{\tau''}{\tau'_{{\rm r},l}} \frac{\tau''}{\tau'_{{\rm r},m}}
\frac{ \langle|\Delta \tilde P_{\rm tot}|^2\rangle +
\langle|\tilde \Gamma_{\rm r} |^2\rangle +
\langle|\tilde \Gamma_{\rm nr}|^2\rangle}
{1+ (\Omega \tau'')^2}
\\
&+& \frac{\tau''}{\tau'_{{\rm r},l}} 
\frac{\langle \tilde \Gamma_{\rm r}^{*} \tilde F_{{\rm r},m}  \rangle}
{1-i \Omega \tau''} +
\frac{\tau''}{\tau'_{{\rm r},m}}
\frac{\langle \tilde \Gamma_{\rm r}^{*} \tilde F_{{\rm r},l}  \rangle}
{1+i \Omega \tau''} 
+\langle \tilde F_{{\rm r},l}^{*} \tilde F_{{\rm r},m} \rangle,
\end{eqnarray}
where 
a total pump noise $\Delta \tilde P_{\rm tot}$ has been defined as
\[
\Delta \tilde P_{\rm tot}(\Omega) 
\equiv \Delta \tilde P (\Omega) + \tilde \Gamma_{\rm P} (\Omega),
\]
which consists of the modulation $\Delta \tilde P (\Omega)$ and
the intrinsic pump noise $\tilde \Gamma_{\rm P} (\Omega)$.
Note that the latter noise can be suppressed by inelastic scattering in 
conductors \cite{YM,TRS,HK,SHKY,Shmz}. 

To see the physical meaning of Eq.\ (\ref{out}), 
let us introduce the Fano factor of 
the pump {\it e}lectrons (or {\it e}xcitons) $ W_{\rm e}$
and that of the {\it ph}otons detected at the PD surface $W_{\rm ph}$,
which are defined by
\begin{eqnarray}
  \label{fano1}
  W_{\rm e}(\Omega) &\equiv& 
        \frac{\langle|\Delta \tilde P_{\rm tot}(\Omega)|^2\rangle}{P_0 T},
\\
  \label{fano2}
  W_{\rm ph}(\Omega) &\equiv& 
        \frac{\langle|\Delta \tilde N(\Omega)|^2 \rangle}{N_0 T},
\end{eqnarray}
where $T$ is the Fourier-integral time. 
We transform Eqs.\ (\ref{cgr})-(\ref{cnrp}) into the Fourier components:
\begin{eqnarray}
\nonumber
&&\langle \tilde F_{\rm r}^*(\Omega) \tilde F_{\rm r}(\Omega) \rangle 
=  \langle \tilde \Gamma^*_{\rm r}(\Omega) \tilde \Gamma_{\rm r}(\Omega) 
\rangle 
=\frac{1}{\epsilon_0} \langle \tilde \Gamma^*_{\rm nr}(\Omega) 
\tilde \Gamma_{\rm nr}(\Omega) \rangle 
\\
\label{Fou}
&&=-\langle \tilde \Gamma^*_{\rm r}(\Omega) \tilde F_{\rm r}(\Omega) 
\rangle 
=\frac{n_{{\rm c}0}}{\tau_{{\rm r}0}}T= V_0 T =\frac{P_0 T}{1+ 
\epsilon_0}.
\end{eqnarray}
Substituting Eqs.\ (\ref{out}) and (\ref{Fou}) into (\ref{Ncorr}),
and dividing by $N_0 T$, we finally obtain
\begin{eqnarray}
\label{fano4}
W_{\rm ph}(\Omega)
=
1-\frac{2 \eta_{\rm d} \zeta_1}{1+(\Omega \tau'')^2}
+ \frac{\eta_{\rm d}^2}{\eta} \frac{1+W_{\rm e} (\Omega)}{1+(\Omega 
\tau'')^2} \zeta_2,
\end{eqnarray}
where we have used Eqs.\ (\ref{Nmean}), (\ref{equi}), (\ref{qe})-(\ref{def2}),
and defined the following,
\begin{eqnarray}
\zeta_1
&\equiv&
\sum_{l,m}
\frac{\tau'_{\rm r}}{\tau_{{\rm r},l}'} \frac{\tau_{{\rm r}0}}{(\tau_{{\rm 
r},m})_0}
\frac{\xi_l}{\beta_0} \frac{\xi_m}{\beta_0},
\label{zeta1}
\\
\zeta_2
&\equiv&
\left[ \sum_{l}
\frac{\tau'_{\rm r}}{\tau'_{{\rm r},l}} 
\frac{\xi_l}{\beta_0}
\right]^2,
\label{zeta2}
\end{eqnarray}
which represent effects due to multimodeness of LEDs,
and the consequences are discussed below.
Note also that $\zeta_1 \le 1, \zeta_2 \le 1$.

Equation (\ref{fano4}) 
is our main result which 
gives the Fano factor of the photons detected at the PD surface as 
a function of the Fano factor of the pump,
measuring frequency, and several parameters
($\eta, \eta_{\rm d}, \zeta_1, \zeta_2$, and $\tau''$).
Note that $1/\tau''$ becomes a cutoff frequency of the Fano factor
as well as that of the modulation [see Eq.\ (\ref{mod})].

\section{Discussion}

\subsection{Low-frequency limit}

Let us examine Eq.\ (\ref{fano4}) in two cases, $\Omega=0$ and $\Omega>0$,
separately.
We first discuss the low-frequency limit.
This case is applicable to the most experiments, because 
they are usually performed at the frequency 
which is lower than any other relevant frequencies.
Setting $\Omega \rightarrow 0$, we obtain
\begin{equation}
  \label{fano0}
  W_{\rm ph}(0)
    =1-2 \eta_{\rm d} \zeta_1+\frac{\eta_{\rm d}^2}{\eta} [1+W_{\rm e}(0)] 
\zeta_2.
\end{equation}
There are several cases where this expression becomes simpler:

\subsubsection{The case where photons are
emitted and detected 
homogeneously (the homogeneous case)}
(h-1): When photons in each mode are emitted and detected 
{\it homogeneously},
{\it i.e.}, when $K_{{\rm r},l}= K_{\rm r} =constant$ and $\xi_l = \beta_0 
=constant$,
we have $\zeta_1=\zeta_2=1$, hence
\begin{equation}
\label{fanofs}
W_{\rm ph}(0)= 1-2 \eta_{\rm d} +\frac{\eta_{\rm d}^2}{\eta} [1+W_{\rm e}(0)].
\end{equation}
This is the formula which we have derived elsewhere \cite{FS}.
We have implicitly assumed this property (h-1) there,
and this is a natural choice unless one considers a situation such that
inhomogeneity due to, {\it e.g.}, cavity-QED effects becomes important.
\\
(h-2): In addition to (h-1),
when the non-radiative processes do not exist 
and/or 
the carrier-number dependence of lifetimes cancels to be zero,
{\it i.e.}, $\tau_{{\rm nr}0} \rightarrow \infty$ and/or $K_{\rm 
r}+K_{{\rm nr}}=0$,
we have $\eta=\eta_{\rm d}$ and the I-L (Injection-Light) characteristics 
become straight. Hence, from Eq.\ (\ref{fanofs}), we obtain
\begin{eqnarray}
\label{fanoold}
\nonumber
W_{\rm ph}(0)
&=& 1-2 \eta +\eta [1+W_{\rm e}(0)]
\\
&=& 1-\eta+\eta W_{\rm e}(0).
\end{eqnarray}
This is just the previous formula 
which is frequently used in the literature \cite{TS,Bj,TRS,HK,GKBR}.

\subsubsection{The case where photons are emitted
and/or detected inhomogeneously (the inhomogeneous case)}
Note that in the homogeneous case
the factors $\zeta_1$ and $\zeta_2$ do not
appear in the expressions for $W_{\rm{ph}}$ 
[Eqs.\ (\ref{fanofs}) and (\ref{fanoold})].
On the other hand,
they appear when emission and/or detection efficiencies
are different among different modes.
We call this general case the ``inhomogeneous case."
To investigate this case,
let us consider a simple case where $K_{{\rm r},l}=0$ and $K_{\rm{nr}} \neq 
0$.
In this case, $\zeta_1 = \zeta_2 \equiv \zeta$,
where
\begin{equation}
\zeta
\equiv
\left[
\sum_{l}
\frac{\tau_{{\rm r}0}}{(\tau_{{\rm r},l})_0} \frac{\xi_l}{\beta_0}
\right]^2.
\end{equation}
Hence, from Eq.\ (\ref{fano0}), we obtain
\begin{equation}
\label{fanoih}
W_{\rm{ph}}(0)
=1-2 \eta_{\rm{d}} \zeta+\frac{\eta_{\rm{d}}^2 \zeta}{\eta} 
[1+W_{\rm{e}}(0)].
\end{equation}
Compared with Eq.\ (\ref{fanofs}), we see that
$\eta$ and $\eta_{\rm{d}}$ are effectively multiplied by $\zeta$
in Eq.\ (\ref{fanoih}).
However, this does not mean that $\zeta$ could be absorbed in
$\eta$ and $\eta_{\rm{d}}$, because they are already defined by
Eqs.\ (\ref{qe}) and (\ref{dqe}), respectively.
Any redefinition would lead to disagreement with
the observed I-L characteristics.


\subsubsection{Condition for generation of sub-Poissonian light}

As another illustration of our result, 
we next discuss the condition for generation of sub-Poissonian light (SPL) 
($W_{\rm ph} <1$) with a Poissonian pump ($W_{\rm e}=1$).
For simplicity, we hereafter consider the case (h-1) [Eq.\ (\ref{fanofs})], 
because many LEDs seem to be categorized in this case.
From Eq.\ (\ref{fanofs}), we obtain
\begin{equation}
\label{cond}
0<\eta_{\rm d} < \eta,
\end{equation}
as the condition.
This can be intuitively understood 
that the flatter the I-L characteristics,
the duller the sensitivity of the LED to the pump fluctuation.
Hence,
if I-L characteristics are like Fig.\ 2 (b),
then {\it even a Poissonian pump}
($W_{\rm e}=1$) {\it can produce
sub-Poissonian light}. 
Note that this mechanism is completely different 
from those of Refs.\cite{Bj,GKBR,YN4},
because the authors of \cite{Bj,GKBR,YN4} eventually make 
the current noise injected to the active layers
{\it below Poissonian}. 
On the other hand, 
the condition (\ref{cond}) is equivalent to 
\begin{equation}
K_{\rm r}+K_{{\rm nr}}<0,
\end{equation}
where we have used Eqs.\ (\ref{qe})-(\ref{def2}).
For simplicity, we here take $K_{{\rm nr}}=0$, and 
find below what is needed for $K_{\rm r}<0$.

The spontaneous emission (SE) rate $n_{\rm c}/\tau_{\rm r}$ 
can be expressed approximately as
$n_{\rm c}/\tau_{\rm r} \propto (n_{\rm c})^p$, where $p$ is a constant.
We therefore have $K_{\rm r} \simeq p-1$ from Eq.\ (\ref{tr}).
It is well known \cite{CKS} that $p \simeq 1$
($p \simeq 2$) in high- (low-) injection regions 
for SE processes of free carriers.
For exciton recombination, we have $p \simeq 1$ 
at the LIL.
Thus we usually have non-negative values of $K_{\rm r}$.
Making use of cavity-QED effects,
however, we can obtain {\it negative} values of $K_{\rm r}$.
This is illustrated in Fig.\ 3
for a $p$-doped quantum well structure in a micro cavity,
where we have assumed the following: 
the conduction band is parabolic,
the effective electron mass is 0.1 times 
the free electron mass, 
the doping level is high, and
the cavity-QED effects prohibit SE except at the band edge.
\begin{figure}
\label{fig3}
\caption{
The spontaneous emission (SE) rate as a function of
carrier population in a doped-quantum-well structure
in a micro cavity. 
Three curves correspond to different temperatures:
$T=$ 80K (bottom), $T=$ 15K (middle), and $T=$ 3K (top).
}
\end{figure}
It is seen that
$K_{\rm r}$ becomes negative for the sheet carrier density 
$\gtrsim10^{14}{\rm m}^{-2}$ when temperatures are low enough (maybe below 3K).
In this case, we obtain $\eta_{\rm d}<\eta$
[see Eqs.\ (\ref{qe})-(\ref{def2})], and
even a super-Poissonian pump can produce SPL.
Exciton recombination with cavity-QED effects would work better, 
which will be discussed elsewhere.

It has been therefore shown that our formula (\ref{fano4}) is useful  
to find the condition for generation of SPL, 
or, of course, other quantum states of light.

\subsection{Finite-frequency cases}

Let us turn to the finite frequency cases. 
For simplicity,
we here use Eq.\ (\ref{fano4}) with property (h-1), {\it i.e.}, 
$\zeta_1=\zeta_2=1$,
and assume the case where $K_{\rm r}=K_{{\rm nr}}$ in Figs.\ 4.
\begin{figure}
\label{fig4}
\end{figure}
\begin{figure}
\end{figure}
\begin{figure}
\end{figure}
\begin{figure}
\caption{
The photon Fano factors
are plotted as functions of frequency $\Omega$ 
(in units of $1/\tau_{{\rm r}0}$).
(a) and (b) represent the cases where $K_{\rm r}=K_{{\rm nr}}=0.5$. 
(c) and (d) the cases where $K_{\rm r}=K_{{\rm nr}}=-0.5$. 
The pump noise level is assumed to be 
zero ($W_{\rm e}=0$) and the Poissonian level ($W_{\rm e}=1$) 
in (a), (c) and (b), (d), respectively.
In each figure, the solid and dashed lines represent 
the cases where the non-radiative recombination is significant 
($\tau_{{\rm r}0}/\tau_{{\rm nr}0}=1$) and absent 
($\tau_{{\rm r}0}/\tau_{{\rm nr}0}=0$), respectively.
The cutoff frequencies are indicated by vertical arrows.
}
\end{figure}
The solid lines (dashed lines) in Figs.\ 4 represent the cases
when non-radiative processes exist (do not exist).

\subsubsection{When $K_{\rm r}=K_{{\rm nr}}=0.5$} 
The noiseless ($W_{\rm e}=0$) and Poissonian ($W_{\rm e}=1$) 
pump are denoted by Figs.\ 4 (a) and (b), respectively.
When $\tau_{{\rm nr}0} = \infty$ (dashed lines),
we recover the results similar to those of Yamanishi and Lee \cite{YL}.
We also see that
the cutoff frequency (indicated by vertical arrow) 
for $\tau_{{\rm nr}0} \neq \infty$
becomes higher than that for $\tau_{{\rm nr}0}=\infty$ [Fig.\ 4 (a)].

\subsubsection{When $K_{\rm r}=K_{{\rm nr}}=-0.5$} 
The noiseless ($W_{\rm e}=0$) and Poissonian ($W_{\rm e}=1$) 
pump are denoted by Figs.\ 4 (c) and (d), respectively.
We find that the existence of non-radiative processes gives
smaller $W_{\rm ph}$ in some frequency region [Fig.\ 4 (c)]
or in the entire frequency range [Fig.\ 4 (d)].
In particular, Fig. 4 (d) shows that the
Poissonian pumping can produce SPL in a wide frequency range.
Figure 4 (c) also shows that
the cutoff frequency for $\tau_{{\rm nr}0} \neq \infty$ becomes higher
than that for $\tau_{{\rm nr}0}=\infty$ as well as Fig.\ 4 (a) does.

It might be difficult, however, to observe these features experimentally,
because they would not appear up to $1/\tau_{{\rm r}0} \sim$ 1GHz.
(Recently the photon Fano factor of LEDs has been measured 
up to 40 MHz \cite{SA}.)

\subsection{Comparison with the experimental results}

Hirano and Kuga \cite{HK} reported that the previous formula
(\ref{fanoold})
disagrees with their experimental data in low-frequency regions: 
They measured the following ratio
\begin{equation}
\frac{W_{\rm ph}(W_{\rm e}=0)}{W_{\rm ph}(W_{\rm e}=1)} \equiv r,
\end{equation}
and obtained that $r<1-\eta$, whereas Eq.\ (\ref{fanoold}) gives $r=1-\eta$.
Since the measuring frequency $\Omega_{\rm meas}$ is low enough, {\it 
i.e.}, 
$\Omega_{\rm meas} \sim 10{\rm MHz} \ll 1/\tau'' \sim 1{\rm GHz}$,
and their LEDs seem to be categorized in the homogeneous case (h-1),
we can use Eq.\ (\ref{fanofs}) to analyze their results.
Our formula (\ref{fanofs}) gives
\begin{equation}
\label{r}
r=\frac{1-2 \eta_{\rm d} +\eta_{\rm d}^2/\eta}{1-2 \eta_{\rm d} +2 
\eta_{\rm d}^2/\eta}.
\end{equation} 
When $\eta_{\rm d} >\eta$ (which is usually the case 
in low-injection regions \cite{HK,SHKY,HSAK,SA}),
our theory gives $r<1-\eta$, in agreement with the experimental results.
Furthermore, Hirano {\it et al}.\cite{HSAK} recently
confirmed that Eq.\ (\ref{r}) 
agrees with their experimental data (see the table below).
It is seen that the values of the previous theory ($r=1-\eta$) 
are larger than the
experimental values, whereas the values of Eq.\ (\ref{r}) are closer to them.
It also has been shown that, when $W_{\rm e}=1$,
$W_{\rm ph}$ {\it itself} is more than one \cite{HSAK}.
This fact also supports our formula (\ref{fanofs})
because it gives,
when $W_{\rm e}=1$ and $\eta_{\rm d} >\eta$,
$W_{\rm ph}=1-2 \eta_{\rm d} +2 \eta_{\rm d}^2/\eta >1$,
and their LEDs also have the property $\eta_{\rm d} >\eta$ \cite{HK,HSAK}.

On the other hand,
a simple argument \cite{HSAK,shmz} leads to
\begin{equation}
\label{fanoshmz}
W_{\rm{ph}}(0)= 1-\eta +\frac{\eta_{\rm{d}}^2}{\eta} W_{\rm{e}}(0).
\end{equation}
Since $|\eta - \eta_{\rm{d}}|$ is small in Table I,
the difference between this formula and Eq.\ (\ref{fanofs})
is within the experimental error and is not detectable.
Further experimental studies are needed to observe the difference.

\section{Summary}

From the microscopic QLEs 
(\ref{Langevin_for_sig})-(\ref{Langevin_for_carrier}),
the associated noise correlations (\ref{csig})-(\ref{ce}),
and the assumption (\ref{condition_for_low_injection}),
we have derived  
the effective semiconductor QLEs (\ref{l1}), (\ref{l2})
and the associated noise correlations (\ref{cgr})-(\ref{cnrp}).
The assumption (\ref{condition_for_low_injection}) is valid 
at a low-injection level and in real devices
as explained in appendix A.
Applying the semiconductor QLEs to 
semiconductor LEDs,
we obtain a new formula (\ref{fano4}) for the Fano factor of photons.
It gives the photon-number statistics as a function of the pump statistics
and several parameters of LEDs, which are defined by 
(\ref{cut})-(\ref{def2}), (\ref{zeta1}) and (\ref{zeta2}).
Key ingredients are non-radiative processes,
carrier-number dependence of lifetimes,
and multimodeness of LEDs.
The formula is applicable to the actual cases where the quantum 
efficiency $\eta$ differs from 
the differential quantum efficiency $\eta_{\rm d}$, 
whereas the previous theories \cite{YM,YL,TS,Bj} 
turn out to give $\eta = \eta_{\rm d}$. 
It is also applicable to cases where photons in each mode of the cavity are 
emitted and/or detected inhomogeneously. 
When $\eta_{\rm d} < \eta$ at the running point,
in particular, our formula predicts 
that even a Poissonian pump can produce sub-Poissonian light 
(see Sec.\ IV A 3).
This mechanism for generation of sub-Poissonian light 
is completely different from those of the previous theories, 
which assumed sub-Poissonian statistics for the current injected into the 
active layers of LEDs.
It was shown that our results agree with recent experiments by Hirano, Kuga 
and coworkers \cite{HK,SHKY,HSAK,SA}.
We have also found that, in finite frequency regions,  
non-radiative processes sometimes give
better results (smaller $W_{\rm ph}$ and/or a higher cutoff frequency).
These will deserve further theoretical and experimental researches 
of quantum aspects of light emitted from LEDs.

\section{Acknowledgments}

We would like to thank T.\ Hirano and T.\ Kuga for addressing 
our attention to this problem.
We are also grateful to M.\ Yamanishi, Y.\ Lee, G.\ Shinozaki 
and J.\ Abe for fruitful discussions.
This work has been supported by 
the Core Research for Evolutional Science and Technology (CREST) of 
the Japan Science and Technology Corporation (JST),
and by Grants-in-Aid for Scientific Research on Priority Areas from the
Ministry of Education, Science and Culture.

\begin{appendix}
\section{Validity of the inequality}

We here show that 
the inequality (\ref{condition_for_low_injection}),
$R_{{\rm sp},l}, R_{{\rm abs},l} \ll \kappa_l^0$, holds
for LEDs at a low-injection level (LIL).

At the LIL,
pumped carriers first relax to lower-energy states
which are formed by impurities, defects, spatial randomness and so on,
and then recombine to radiate photons [Fig.\ 5 (a)].
Within the energy region of our interest,
where photons are emitted, 
we thus have $R_{{\rm sp},l} \sim R_{{\rm abs},l}$.
In addition, in real devices, 
there exists a Stokes shift [Fig.\ 5 (b)]
which makes, at low temperatures, $R_{{\rm abs},l}$ even smaller. 
Hence we have the absorption and emission profiles such as Fig.\ 5 (c).

We now proceed to compare $R_{{\rm sp(abs)},l}$ with $\kappa_l^0$.
In Eqs.\ (\ref{Rsp}) and (\ref{Rabs}),
$|g_{l,\bf k}|^2$ is proportional to $1/V_{\rm cavity}$
($V_{\rm cavity}$ is the volume of a cavity), and $\sum_{\bf k}$ is 
to the volume of the active layer $V_{\rm active}$,
hence we have $R_{{\rm sp(abs)},l} \propto V_{\rm active}/V_{\rm cavity}$.
In the analysis of {\it laser diodes} (LDs), 
it is customary to take
$V_{\rm active} \simeq V_{\rm cavity}$, 
because the devices are made so that the lasing modes are confined 
in the active layer.
In this case, we usually have
$R_{{\rm abs},l}/c \sim 10^4 {\rm cm}^{-1} 
\gg \kappa_l^0/c \sim 10^2 {\rm cm}^{-1}$
($c$ is the velocity of light), thus we can not obtain 
the inequality (\ref{condition_for_low_injection}).
In the case of {\it LEDs}, on the other hand,
they are usually designed in such a way
that the reflection coefficients on the 
boundary surfaces are small, 
hence most modes of photons of our interest 
are not confined in the layer.
In this case, it is natural to take 
the ``cavity'' volume as big as a cube on which the
detector's surface is located [Fig.\ 5 (d)].
Then $\kappa_l^0$ can be estimated as 
$1/\kappa_l^0 \simeq (V_{\rm cavity})^{1/3}/c + Q t_{\rm device}$,
where $t_{\rm device}$ is a time for photons to traverse the device.
Since $Q$ and $t_{\rm device}$ are small for LEDs,
we have $\kappa_l^0 \simeq c/(V_{\rm cavity})^{1/3}$.
Therefore, 
noting that $R_{{\rm abs},l} \propto 1/V_{\rm cavity}$ and 
$\kappa_l^0 \propto 1/(V_{\rm cavity})^{1/3}$,
we can have the relation $R_{{\rm abs},l} \ll \kappa_l^0$,
if we take $V_{\rm cavity}$ big enough.
Thus we obtain the inequality (\ref{condition_for_low_injection})
for the case of LEDs at the LIL.
\begin{figure}
\label{fig5}
\caption{
(a) A schematic band diagram of real LEDs at the LIL, 
(b) the Stokes shift
[CB (VB) represents the conduction (valence) band.], 
(c) the absorption and emission profiles at the LIL,
and
(d) the active layer, the device (LED), and the ``cavity". 
}
\end{figure}
\end{appendix}


\begin{table}
\caption{The experimental values of $\eta, \eta_{\rm d}, r$ 
and the theoretical values of $r$. After Ref. [11].}
\begin{tabular}{c|c|c|c|c} \hline
 $\eta$(exp.)& $\eta_{\rm d}$(exp.) & $r$(exp.) 
 & $r$(Eq.\ (\ref{r})) & $r=1-\eta$ \\ \hline
0.067 & 0.090 & 0.90 & 0.89 & 0.93  \\ \hline
0.104 & 0.125 & 0.84 & 0.86 & 0.90  \\ \hline
0.150 & 0.175 & 0.81 & 0.81 & 0.85  \\ \hline
\end{tabular}
\end{table}
%
%
%
%
%
%
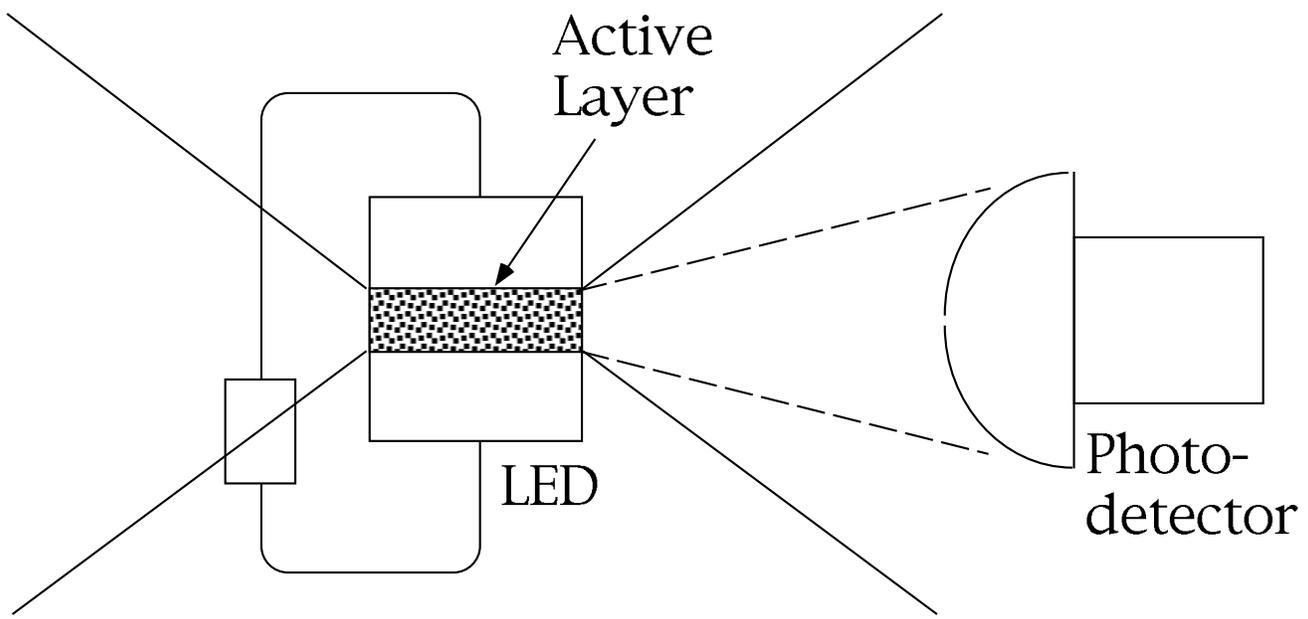
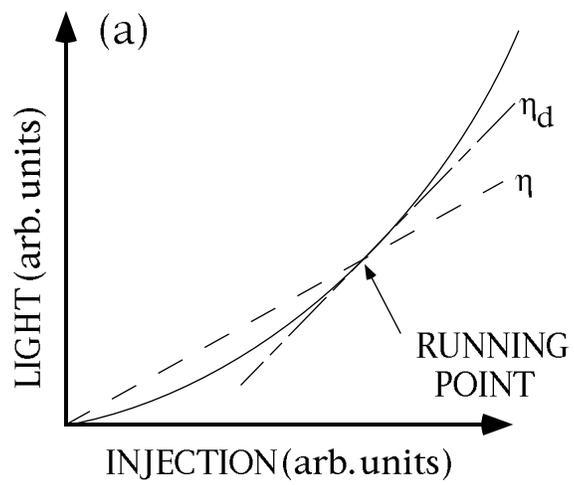
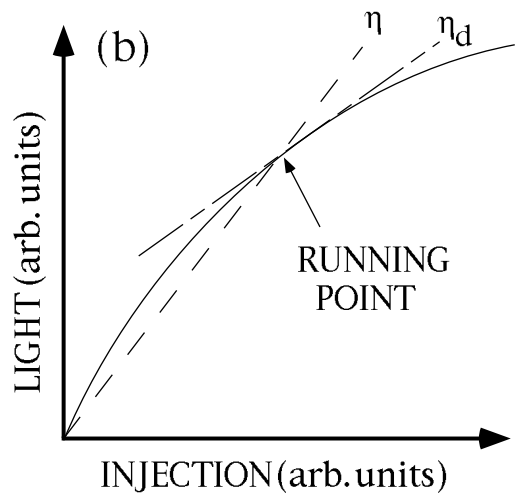
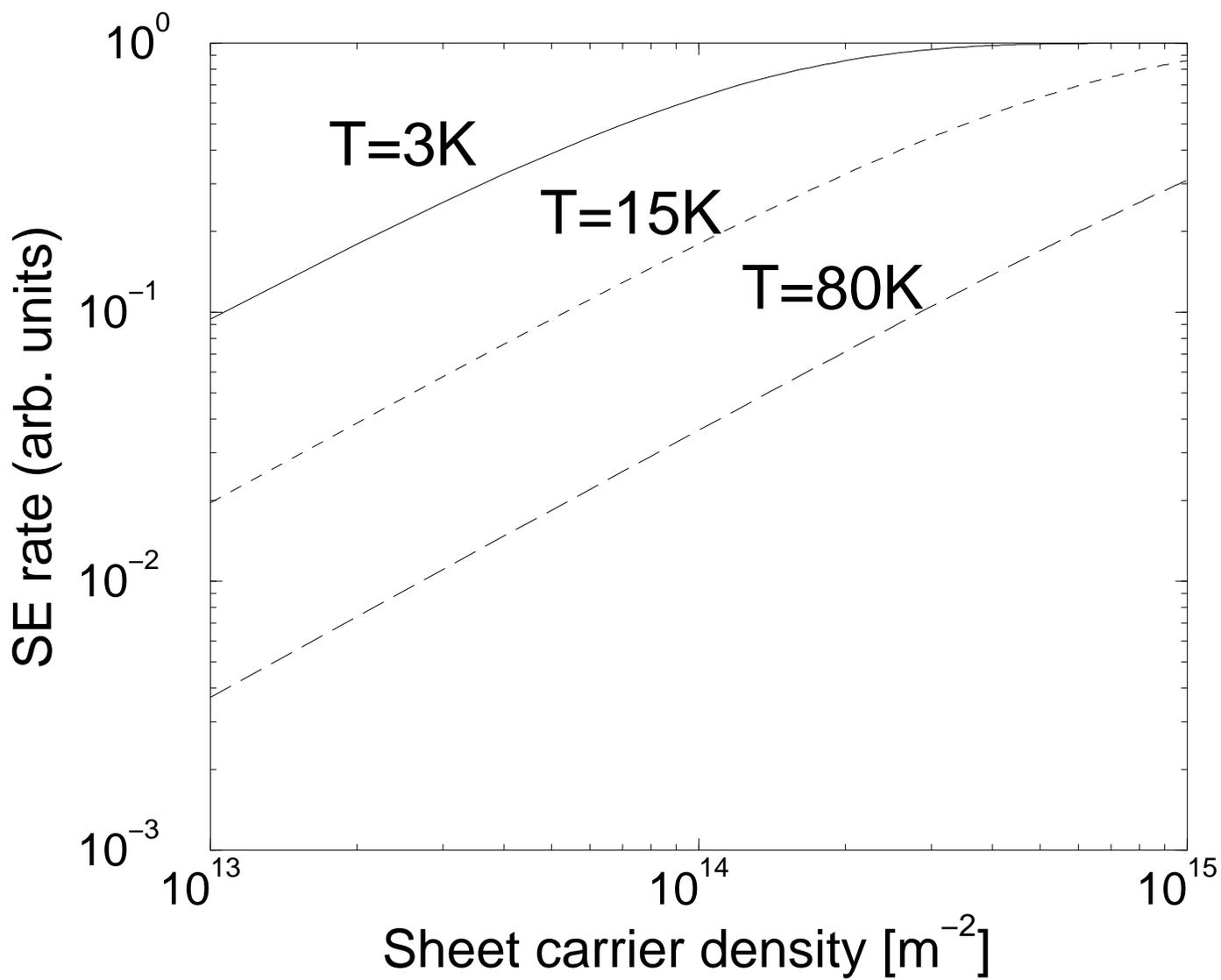
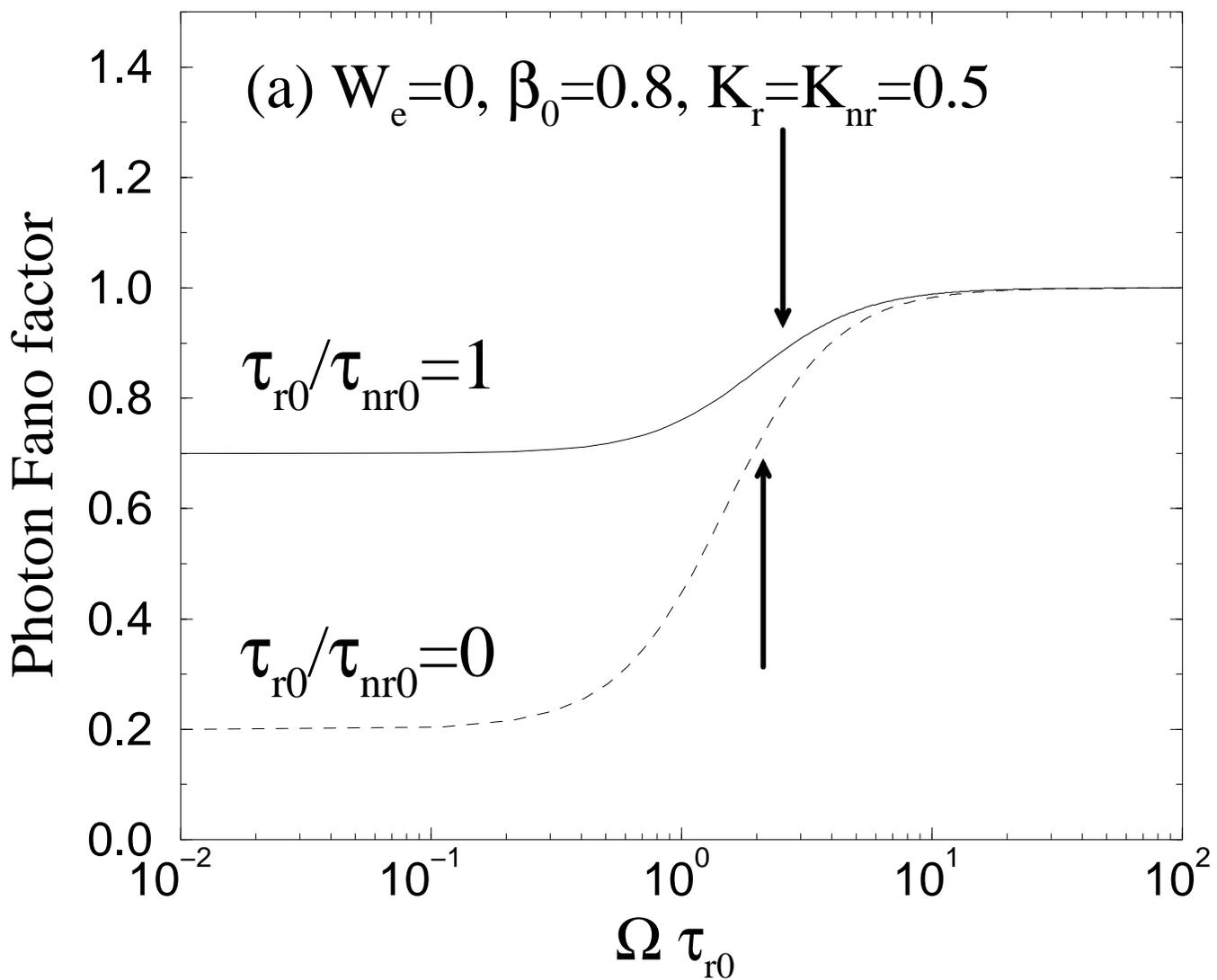
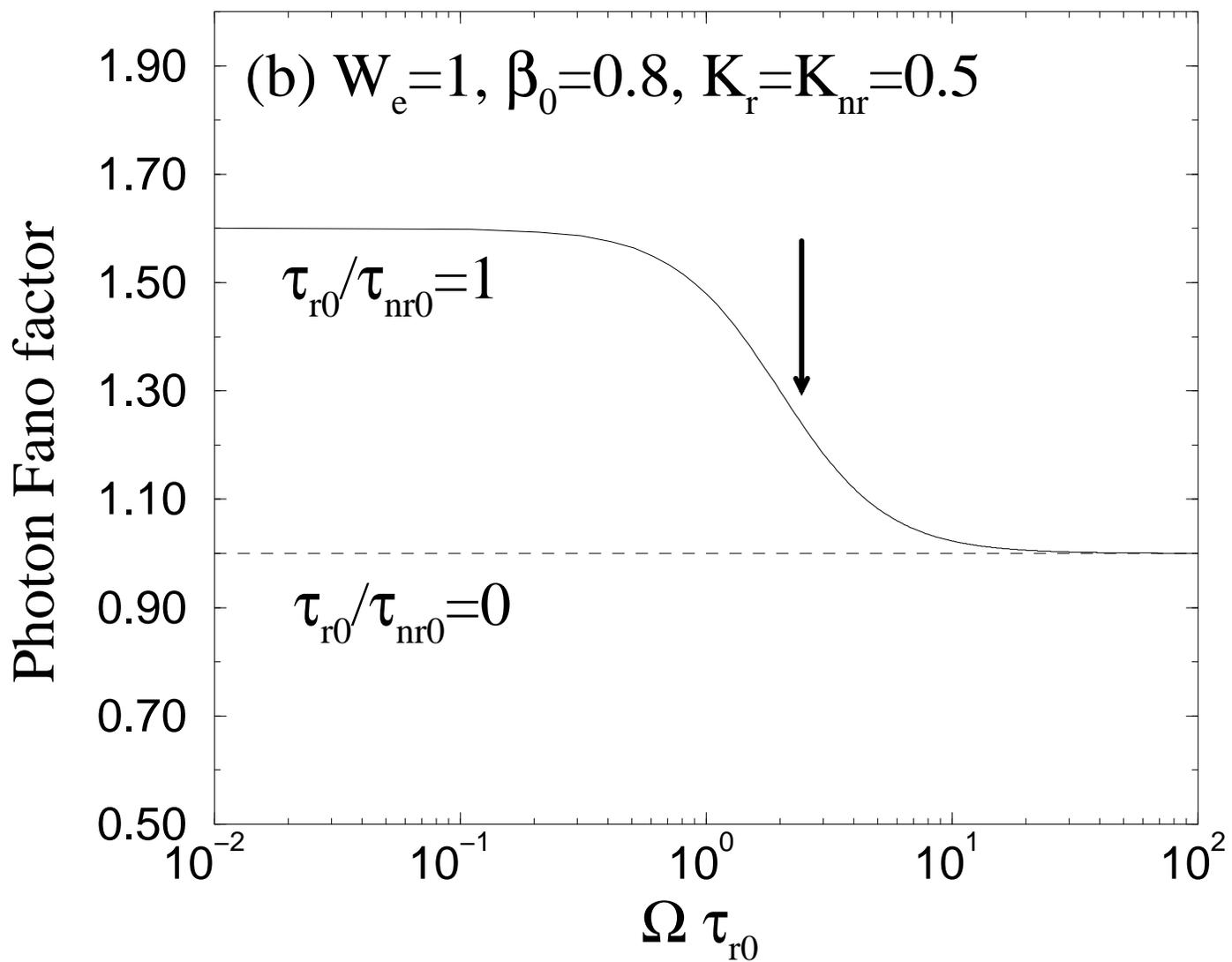
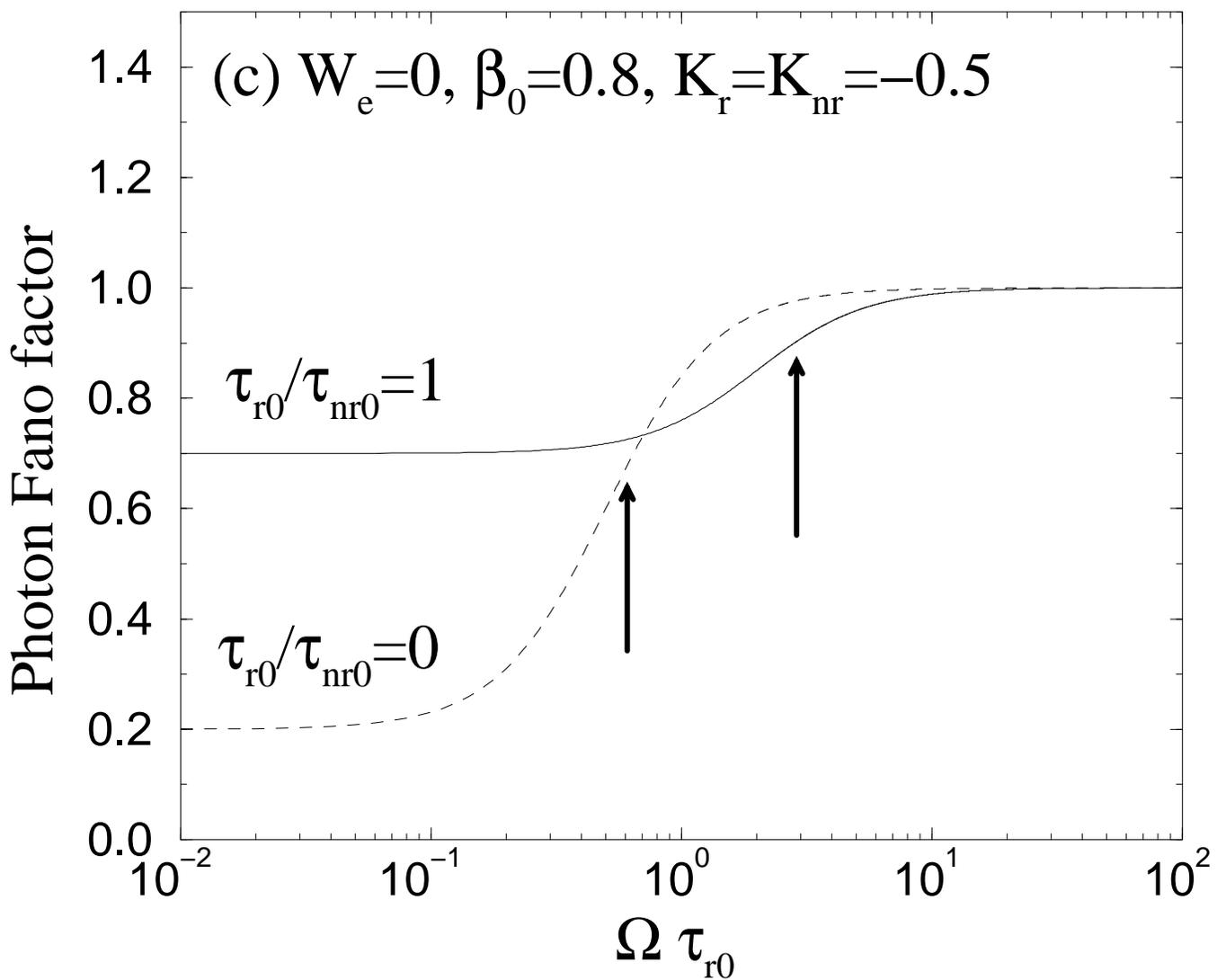
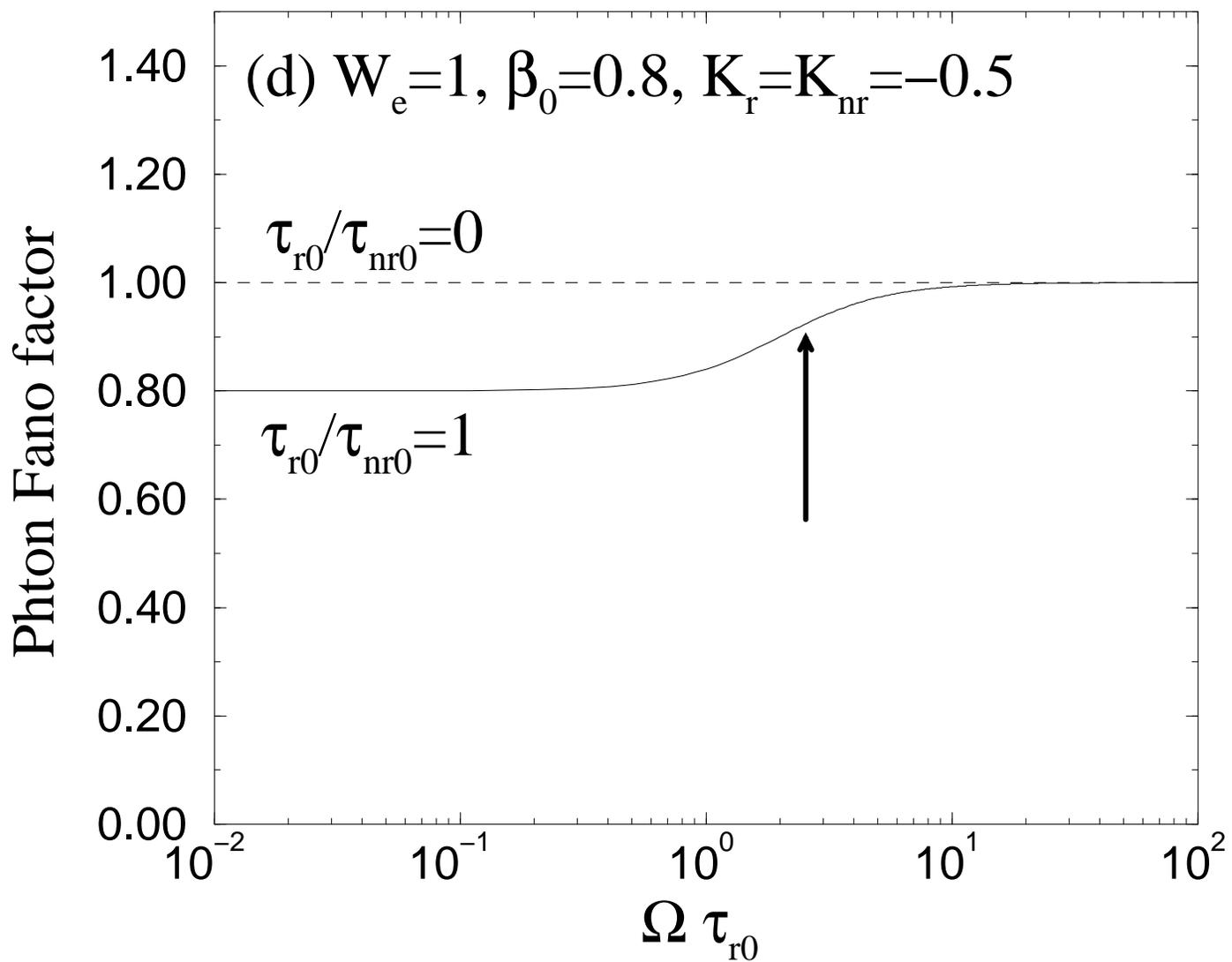
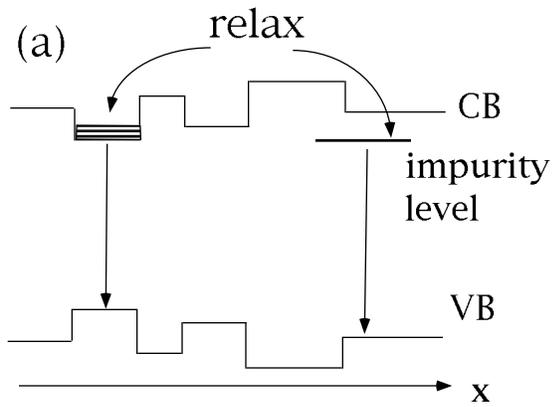
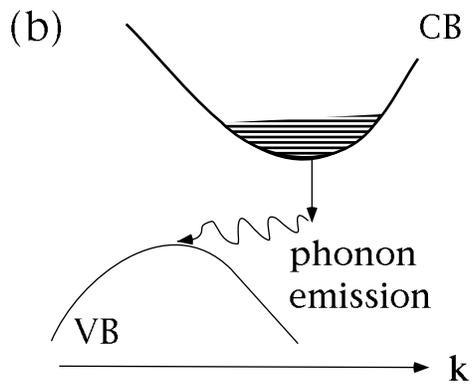
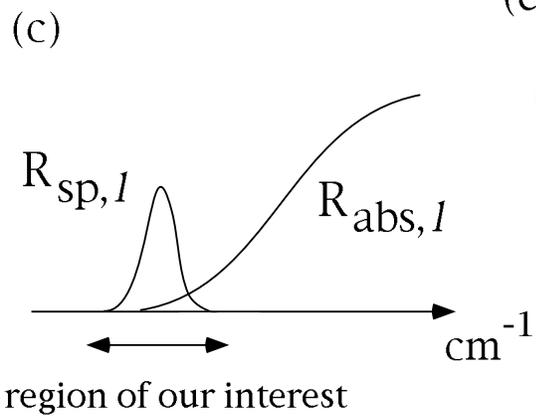
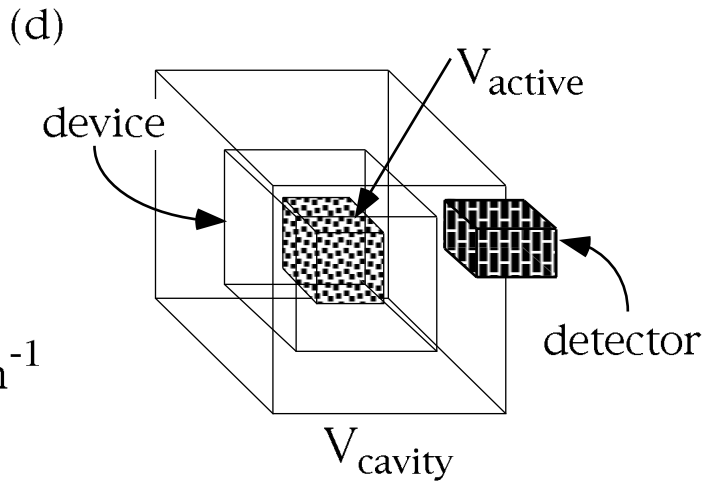
\end{document}